\pdfoutput=1
\documentclass[11pt,a4paper]{article}

\usepackage[T1]{fontenc}
\usepackage[utf8]{inputenc}
\usepackage[margin=1in]{geometry}
\usepackage{microtype}
\usepackage{booktabs}
\usepackage{array}
\usepackage{amsmath}
\usepackage{listings}
\usepackage{xcolor}
\usepackage{tikz}
\usetikzlibrary{arrows.meta,positioning,fit,backgrounds,calc}
\usepackage[hidelinks,breaklinks]{hyperref}
\usepackage{cleveref}

\definecolor{codebg}{gray}{0.96}
\lstdefinestyle{kah}{
  basicstyle=\ttfamily\small,
  backgroundcolor=\color{codebg},
  breaklines=true,
  columns=fullflexible,
  frame=none,
  xleftmargin=1em,
  showstringspaces=false,
  literate={–}{{-}}1 {—}{{---}}1 {“}{{``}}1 {”}{{''}}1 {’}{{'}}1
}
\tikzset{
  box/.style   = {draw, rounded corners=2pt, align=center, inner sep=5pt,
                  font=\small},
  op/.style    = {box, fill=black!4},
  sink/.style  = {box, fill=black!8, very thick},
  lbl/.style   = {font=\scriptsize\itshape, align=center},
  ar/.style    = {-{Stealth[length=5pt]}, thick},
  bnd/.style   = {draw, dashed, rounded corners=4pt, inner sep=11pt}
}

\newcommand{\tier}[1]{\texttt{#1}}

\title{\textbf{Separating Disclosure from Authorization:}\\[2pt]
       \large Field-Tier Minimization for Agent Action Mediation}

\author{%
  Jiten Oswal \qquad John Cadeddu\\[5pt]
  \normalsize Aurite AI\\[3pt]
  \normalsize \texttt{jiten@aurite.ai} \qquad \texttt{john@aurite.ai}\\[1pt]
  \normalsize \footnotesize ORCID (J.~Oswal):
    \href{https://orcid.org/0009-0009-8866-869X}{0009-0009-8866-869X}%
}

\date{August 25, 2026}

\begin{document}
\maketitle

\begin{abstract}
\noindent
A system that authorizes an action must see enough of it to decide, and a system
that attests to its decision must record enough to be audited. Both pressures
push raw action parameters --- recipients, payment memos, record identifiers ---
into an append-only ledger that cannot delete them. We show the two are
separable. We classify each parameter \emph{field}, not each action class, into
three tiers: fields a policy may legitimately match on, which cross raw; fields
that are policy-relevant but identifying, which cross only as projections such
as an email domain or a templated route shape; and fields with no legitimate
policy use, which never leave the workload. The central property is that the
ledger's commitment is a canonical digest of the \emph{full, unminimized}
parameters, computed \emph{before} minimization runs. The commitment is
therefore independent of the tier table: reclassifying a field changes what is
disclosed without invalidating a historical entry, reopening a hash, or altering
what an offline verifier checks. Tier table, policy schema and wire schema are
generated from one per-action declaration, so the deciding and recording parties
cannot hold different rules. We then address a question the architecture forces:
\emph{which party should compute each attested fact?} We argue it is settled by
which party could lie about it undetectably, and derive three answers within one
request --- the client computes the parameter digest, being the only party
holding the data; it is structurally prevented from naming the definition that
governed it, since that would write a false statement into a signed ledger; and
it attests which tier table it applied, so divergence is detectable. We give a
leakage analysis of each projection, report an incident in which a first-cut
projection preserved the identifier it was written to remove, and state the
residual trust the design does not eliminate.
\end{abstract}

\section{Introduction}

Consider an autonomous agent that sends an email on a customer's behalf. To
authorize it, a policy engine must evaluate a rule --- \emph{may this agent
email outside the company?} --- which requires knowing something about the
recipient. To audit it, the system must record what happened, durably, in a form
an auditor can check months later. Both requirements are legitimate. Both,
implemented naively, place the recipient address, the subject line, and the body
in a mediation service and then in an append-only log.

That log is the problem. An audit ledger built for tamper-evidence is
deliberately hard to change: rows are hash-linked and signed, deletion breaks
the chain, and retention is measured in years~\cite{crosby2009,rfc6962}.
Whatever enters it is there for the life of the record. A system that governs
agents therefore concentrates, in its most immutable component, precisely the
data its customers are most regulated about~\cite{gdpr}, and does so as a
\emph{side effect} of doing its job well.

The standard responses are unsatisfying. Redacting at write time means the
ledger commits to something other than what happened, and an auditor cannot tell
what was removed. Hashing everything means no policy can match on anything.
Encrypting the payload moves the problem to key management and leaves the
plaintext recoverable by whoever holds the key, which for an audit ledger is
the party whose behaviour is under audit.

\paragraph{Our claim.} The two pressures are separable, and the separation is
cheap if the commitment is placed correctly. The policy engine needs
\emph{predicates over} parameters, not parameters. The ledger needs a
\emph{commitment to} parameters, not parameters. Neither needs the values
themselves to cross the boundary, and the values that must cross for policy
reasons need not be the identifying ones.

This paper describes a design that exploits this, implemented and running in a
runtime governance system for enterprise AI agents. That system's overall
architecture --- and the argument for why governing agents is a runtime problem
rather than a build-time or alignment one --- is described
separately~\cite{c1primitives}; minimization is one mechanism inside its
governance layer, and this paper is self-contained with respect to it. The
system is operated as a small number of private pilots rather than as a general
release. That is a deliberate choice we return to in \cref{sec:scope}, because
it determines what this paper can and cannot claim.

\paragraph{Contributions.}
\begin{enumerate}\itemsep2pt
  \item \textbf{Field-level, not class-level, minimization} (\cref{sec:tiers}),
        with a three-tier vocabulary and a declassifying projection library.
  \item \textbf{Digest independence} (\cref{sec:digest}): the ledger commits
        to a digest of the raw parameters computed before minimization, which
        makes the disclosure policy \emph{retrofittable and re-tierable} over an
        immutable ledger.
  \item \textbf{One declaration, three artifacts} (\cref{sec:decl}): the tier
        table, the policy schema, and the wire schema are generated from one
        per-action declaration, so the deciding party and the recording party
        cannot disagree.
  \item \textbf{An asymmetry principle for attested facts}
        (\cref{sec:asymmetry}): \emph{the party that computes an attested fact
        should be the party that cannot lie about it undetectably.} This yields
        three different answers within a single request, and getting one wrong
        produces a false statement in a signed ledger, a failure strictly worse
        than a wrong authorization decision.
  \item \textbf{Disclosure is not authorization} (\cref{sec:disclosure}):
        deny-by-default protects one and not the other.
  \item \textbf{A leakage analysis} (\cref{sec:leak}) of each projection:
        what each still reveals, a quantification over our corpus, and an
        incident in which a first-cut projection preserved the identifier it was
        written to remove.
\end{enumerate}

\Cref{app:worked} traces one request end to end (raw parameters, digest,
projected attributes, policy evaluation, attested event) using values pinned as
literals in our test suite. A reader who wants the concrete artifact before
the argument should start there.

We do not claim novelty for tiered data classification, nor for canonical
hashing. The contribution is the \emph{placement of the commitment}, the
asymmetry principle it forces, and the demonstration that together they make
disclosure policy mutable over an immutable record.

\section{Setting}\label{sec:setting}

\subsection{The mediation path}

An agent attempting a governed action calls an authorization endpoint before the
action takes effect. The mediator evaluates a policy (in our implementation,
Cedar~\cite{cedar}) against an entity model in which the agent is the
principal, the action type is the action, and the action's parameters supply
\emph{resource attributes}. The decision is returned to the agent, which
proceeds or does not, and the decision is written to an append-only,
hash-linked, signed ledger.

The parameters are therefore load-bearing twice: once as policy inputs, once as
the audited record of what was attempted.

\subsection{Trust boundaries}\label{sec:boundaries}

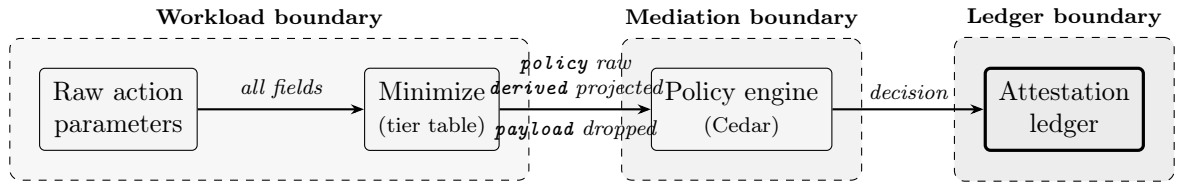
\begin{figure}[t]
\centering
\begin{tikzpicture}[node distance=6mm]
  \node[op] (params) {Raw action\\parameters};
  \node[op, right=22mm of params] (min) {Minimize\\\scriptsize(tier table)};
  \node[op, right=20mm of min]    (cedar) {Policy engine\\\scriptsize(Cedar)};
  \node[sink, right=20mm of cedar] (ledger) {Attestation\\ledger};

  \draw[ar] (params) -- node[lbl, above] {all fields} (min);
  \draw[ar] (min) -- node[lbl, above] {\tier{policy} raw\\\tier{derived} projected}
        node[lbl, below] {\tier{payload} dropped} (cedar);
  \draw[ar] (cedar) -- node[lbl, above] {decision} (ledger);

  \begin{scope}[on background layer]
    \node[bnd, fill=black!3, fit=(params)(min), label={[font=\scriptsize\bfseries]above:Workload boundary}] (b1) {};
    \node[bnd, fill=black!5, fit=(cedar), label={[font=\scriptsize\bfseries]above:Mediation boundary}] (b2) {};
    \node[bnd, fill=black!7, fit=(ledger), label={[font=\scriptsize\bfseries]above:Ledger boundary}] (b3) {};
  \end{scope}
\end{tikzpicture}
\caption{The three boundaries. Exposure at the mediation boundary is bounded in
time and audience; exposure at the ledger boundary is neither, because the
record is permanent and designed to be exported to third parties. Treating them
as one boundary is the error this design avoids.}
\label{fig:boundaries}
\end{figure}

Three boundaries matter, and conflating them is the mistake this design exists
to avoid (\cref{fig:boundaries}).

\paragraph{The workload boundary.} The agent runs in the customer's
environment. Code we ship runs inside it, as an SDK. Data that never leaves the
process has not crossed any boundary at all. That is the strongest position
available, and the one the \tier{payload} tier occupies.

\paragraph{The mediation boundary.} The policy engine may run in the customer's
account or ours. Either way it is a different failure domain from the workload,
and anything crossing it is disclosed in the sense that matters: it exists
somewhere the workload does not control.

\paragraph{The ledger boundary.} Anything written here is effectively permanent
and is designed to be exportable to third parties, which is the point of an
evidence package. Data crossing this boundary is disclosed to every future
auditor.

A design that treats ``the mediator is trusted'' as equivalent to ``the ledger
may hold anything'' has collapsed the second and third boundaries, which are not
the same boundary.

\subsection{What ``minimization'' must not break}\label{sec:constraints}

Three properties were non-negotiable, and they constrain the design more than
the privacy goal does.

\begin{itemize}\itemsep2pt
  \item \textbf{The attested record must commit to what actually happened}, not
        to a redacted view of it. An auditor must be able to establish that a
        specific parameter set produced a specific decision.
  \item \textbf{Existing chain entries must remain verifiable}, byte-for-byte,
        under an offline verifier that has already been published for
        independent use.
  \item \textbf{The policy language must not become weaker in ways operators
        cannot see.} If a policy can no longer express something, that must be
        visible at authoring time, not as a silent non-match at runtime.
\end{itemize}

\section{Field-tier minimization}\label{sec:tiers}

\subsection{The unit is the field}\label{sec:field-not-class}

The natural first design classifies \emph{action types}: \texttt{send\_email} is
sensitive, \texttt{read\_data} is not. This fails immediately on inspection of
any real action. \texttt{send\_email} carries a recipient (identifying, and
policy-relevant), a subject (identifying, and in our corpus referenced by zero
policies), an attachment count (not identifying, and
policy-relevant), and a body (never policy-relevant under any rule we have
seen). A per-class decision must either disclose the subject to permit a
recipient rule, or refuse recipient rules to protect the subject. Both are
wrong.

So the unit of classification is the field, classified once, statically, per
action type.

\subsection{Three tiers}

\begin{table}[h]
\centering\small
\begin{tabular}{@{}lp{63mm}l@{}}
\toprule
\textbf{Tier} & \textbf{Rule} & \textbf{Crosses as} \\
\midrule
\tier{policy}  & A policy may legitimately match on it, and it does not identify a person or record & The raw value \\
\tier{derived} & Policy-relevant, but the raw value identifies & A non-identifying projection \\
\tier{payload} & No legitimate policy use & Nothing --- folded only into the digest \\
\bottomrule
\end{tabular}
\end{table}

The \tier{derived} tier is the interesting one and the reason field-level
classification is worth its cost. It exists because the \emph{policy-relevant}
part of an identifying value is usually a strict, computable abstraction of it.
\emph{``May this agent email outside the company?''} is a question about a
domain; the local part identifies a person and answers nothing.

In information-flow terms~\cite{denning1976,sabelfeld2003} the tiers are a
two-point lattice with an explicit declassification set: \tier{payload} is high,
\tier{policy} is low, and each \tier{derived} field carries a named declassifier
mapping high to low. The design is a \emph{declassification
policy}~\cite{sabelfeld2009}, and much of what follows, particularly
\cref{sec:applied-tiers}, is about making that policy itself auditable.

\subsection{The projections}\label{sec:projections}

Five projections cover our corpus. Each is pure, individually unit-tested, and
compiled into every client.

\begin{description}\itemsep3pt
\item[\texttt{recipient\_domain}] The substring after the last \texttt{@},
  lowercased. A recipient with no \texttt{@} has no domain and yields
  \texttt{""}, which matches no domain policy. Note the direction of that
  failure: an unparseable input fails to a value that \emph{denies} under
  allowlist policies rather than to the raw value.

\item[\texttt{payee\_known}] A boolean: is this payee in the operator's
  allowlist? The payee itself never crosses. An absent allowlist is treated as
  empty, so every payee is unknown, which again fails closed.

\item[\texttt{endpoint\_host}] The lowercased host of an external endpoint. The
  implementation prefers the WHATWG URL parser but only trusts it when it yields
  a real authority, because \texttt{new URL("host:port/p")} parses \texttt{host}
  as a \emph{scheme} with an empty hostname; scheme-less service identifiers
  fall back to an authority/path split.

\item[\texttt{endpoint\_path\_template}] The route \emph{shape}: the path with
  concrete instance identifiers replaced by \texttt{\{id\}} and the query string
  dropped. Numeric segments, UUIDs and long hex tokens are treated as
  identifiers; version-like segments (\texttt{v1}) and named routes
  (\texttt{patients}, \texttt{records}) are preserved, because they are the
  route's meaning rather than its subject. Thus
  \begin{center}\ttfamily\small
  https://api.x.com/v1/patients/40b1.../records?token=abc\\
  $\rightarrow$ /v1/patients/\{id\}/records
  \end{center}

\item[\texttt{resource\_path\_prefix}] The \emph{directory} a resource lives in,
  with identifier-like segments templated and the final segment removed
  entirely. This is the projection that got it wrong the first time, and
  \cref{sec:incident} is about that.
\end{description}

\subsection{Two safety rules that make incompleteness harmless}

\paragraph{Unclassified means \tier{payload}.} A field a definition does not
mention is dropped, not passed. There is deliberately no ``default tier''
parameter anywhere in the API. This is what makes a half-finished action
definition leak nothing: the failure mode of forgetting to classify a field is
that a policy about it silently does not match --- visible to the operator,
fixable --- rather than that the field is disclosed, which is neither.

\paragraph{An unknown projection is refused, not approximated.} A definition
naming a projection the client build does not have is rejected \emph{for that
field}. Applying the closest available projection instead would substitute one
disclosure decision for another without anyone deciding to.

Both rules follow the same principle: when the system is unsure what a field
is, the safe answer is the one that reduces disclosure, even at the cost of
function. This is the mirror image of deny-by-default, applied to a
dimension deny-by-default does not cover (\cref{sec:disclosure}).

\section{Digest independence}\label{sec:digest}

\subsection{The property}

\begin{figure}[t]
\centering
\begin{tikzpicture}[node distance=8mm]
  \node[op] (p) {Raw parameters};

  \node[op, above right=5mm and 20mm of p] (hash) {$\mathrm{SHA\text{-}256}(\mathrm{JCS}(\cdot))$};
  \node[sink, right=20mm of hash] (led) {Attestation ledger\\\scriptsize commitment};

  \node[op, below right=5mm and 20mm of p] (mini) {Minimize\\\scriptsize(tier table)};
  \node[sink, right=20mm of mini] (ced) {Policy engine\\\scriptsize attributes};

  \draw[ar] (p) -- (hash);
  \draw[ar] (p) -- (mini);
  \draw[ar] (hash) -- (led);
  \draw[ar] (mini) -- (ced);

  \node[lbl, right=2mm of led.south east, anchor=north east, yshift=-3mm]
       {independent of the tier table};
  \node[lbl, right=2mm of ced.south east, anchor=north east, yshift=-3mm]
       {changes when tiers change};
\end{tikzpicture}
\caption{The fork. Both computations take the raw parameters and share nothing
else. Because the commitment is taken \emph{before} minimization, reclassifying
a field moves the lower path only --- no historical hash is reopened and the
offline verifier is untouched.}
\label{fig:fork}
\end{figure}
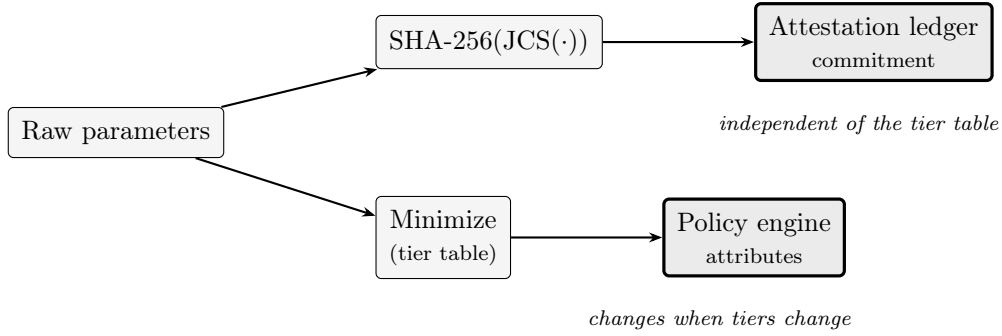

The attested event records \texttt{action\_parameters\_hash} and
\texttt{action\_parameters\_size\_bytes}. Both are computed by the client, over
the full original parameters, \emph{before} minimization runs:
\[
  \texttt{action\_parameters\_hash} \;=\; \mathrm{SHA\text{-}256}\bigl(\mathrm{JCS}(\textit{parameters})\bigr)
\]
where JCS is RFC 8785 canonical JSON serialization~\cite{rfc8785} and SHA-256 is
as specified in FIPS 180-4~\cite{fips1804}. Minimization then runs separately,
producing the Cedar attributes that cross the boundary. The two computations
share an input and nothing else (\cref{fig:fork}).

The consequence is the central property of this design:

\begin{quote}
\textbf{The ledger's commitment does not depend on the tier table.}
\end{quote}

Reclassifying a field --- moving \texttt{subject} from \tier{payload} to
\tier{policy}, or narrowing \texttt{resource\_path} from a full path to a
directory --- changes what crosses the boundary and changes nothing about what
the ledger committed to.

\subsection{Why this matters more than it first appears}

An immutable ledger and a mutable disclosure policy are, on their face,
incompatible. If the commitment were computed over the \emph{minimized} view,
then every tier change would fork the meaning of the hash: entries before and
after would commit to differently-shaped objects, an offline verifier would need
the tier table as of each entry's date to interpret it, and the tier table would
itself become an input to verification, and therefore something an attacker
could attack.

Placing the commitment before minimization avoids all of that:

\begin{itemize}\itemsep2pt
  \item \textbf{Tier changes are purely additive.} No historical entry's hash
        changes.
  \item \textbf{The offline verifier is untouched.} It already shipped; it
        verifies signatures and chain linkage over bytes, and those bytes do not
        move.
  \item \textbf{There is no version-dependent hash to get wrong.} The design is
        \emph{attach a pointer}, never \emph{redefine a commitment}.
  \item \textbf{Minimization is retrofittable.} We added it to a system with a
        populated chain, and entries written before it remain valid, comparable
        and verifiable against entries written after.
\end{itemize}

The last point is the practical one. A design that required minimization from
day zero would be a design nobody could adopt.

\subsection{What the commitment does and does not prove}

The ledger entry proves: \emph{a client holding parameters whose canonical
digest is $H$ requested action $A$, and the mediator decided $D$.} An auditor in
possession of a candidate parameter set can confirm it is the one: the digest is
over the raw values, so the check is exact and requires no knowledge of the tier
table.

It does not prove that the mediator \emph{saw} those parameters; by construction
it did not. \Cref{sec:residual} is about the residual trust that follows.

\section{One declaration, three artifacts}\label{sec:decl}

\subsection{The drift problem}\label{sec:drift}

The tier table is not the only artifact describing an action. Three exist: the
policy schema (the Cedar entity types and attribute names a policy may
reference), the tier table, and the wire schema (runtime validation of an
incoming request).

Written independently, they drift, and each drift is a distinct failure. A field
in the policy schema but not the tier table is a policy that can never match. A
field in the tier table but not the wire schema is an attribute accepted without
validation. Worst, a field classified \tier{policy} in the server's copy and
\tier{payload} in a client's copy means the two parties disagree about what may
be disclosed, and the disagreement is invisible, because each behaves correctly
by its own table.

That last case is not hypothetical. Our Python SDK carried a duplicated tier
table, and it drifted.

\subsection{Generation from one declaration}\label{sec:generation}

All three artifacts are now generated from a single per-action declaration: each
field's tier, its schema, its Cedar type, and, for \tier{derived} fields, the
projections it maps to. The server and the primary client SDK import that
declaration rather than restating it.

Our second-language SDK is the honest qualification. It implements the tier
vocabulary and the digest, and cross-language conformance is enforced by tests
comparing digests produced by both implementations over a shared corpus, so
digest divergence fails in CI. It does \emph{not} yet implement the runtime
definition distribution or the applied-tiers attestation of
\cref{sec:applied-tiers}. The result is that our strongest guarantee --- that
the deciding and recording parties cannot hold different disclosure rules ---
currently holds for one client language and is enforced by conformance testing
rather than by construction for the other. We state this because a generation
argument that quietly excludes a shipped client is the kind of claim this paper
is otherwise arguing against.

\subsection{An aside on testing generators, which we got wrong first}\label{sec:golden}

The generator's first safety argument was an \emph{equivalence} test: assert the
generated tier table matches the hand-written one. That test was correct, and it
became worthless the moment the hand-written table was deleted --- at which
point it compared the generator to itself. Forty-nine assertions, all vacuously
true, sitting in CI looking like coverage. The failure is the operational
analogue of vacuous satisfaction in model checking~\cite{beer2001,kupferman2003},
where a property passes because its antecedent never occurs.

The replacement is a golden corpus of literals: every expected value
written out by hand, captured from the reviewed implementation before the swap.
Literals have independent force precisely because a human wrote them down, and
they keep that force when the generator is itself replaced. The relationship to
mutation testing~\cite{demillo1978,jia2011} is close: what makes a check
meaningful is evidence that it can fail.

The leak guard in that corpus is worth describing, because the obvious version
of it does not work. It does not check that identifying values are absent from
the keys they belong to; it serializes the entire minimized output and asserts
the identifying values appear nowhere in it, under any key. The failure being
guarded is a projection that returns its input, which a per-key assertion would
miss entirely, since the value would appear exactly where a correct
implementation puts a correct value.

\section{Who computes what: an asymmetry principle}\label{sec:asymmetry}

Minimization forces a question that a non-minimizing mediator never has to ask.
If the server does not see the raw parameters, it cannot compute facts about
them, so some attested facts must be computed by the client. Which ones?

\begin{quote}
\textbf{The party that computes an attested fact should be the party that cannot
lie about it undetectably. Where that is impossible, the fact must not be
attestable by the client at all.}
\end{quote}

Applied to a single authorization request, this yields three different answers.

\subsection{The parameter digest: the client computes it, and is trusted}\label{sec:digest-trust}

Only the client holds the raw parameters. No other party can compute the digest.
The server therefore records what the client reports.

This is a real trust concession and we name it as such: a client that lies
about its own digest produces a ledger entry that commits to parameters other
than the ones it used. Four things bound it.

\paragraph{The claim is not anonymous.} The digest arrives over a
mutually-authenticated channel on which the client presents a short-lived
credential issued to its workload by an attestation-based identity
plane~\cite{spiffe}, on the strength of platform properties it cannot choose for
itself. The ledger therefore does not record \emph{some client asserted $H$}; it
records \emph{this attested workload asserted $H$}.

\paragraph{The party who could forge it is the party it would harm.} The client
is the customer's own workload in the customer's environment, and the ledger
exists to evidence that workload's behaviour to third parties. Forging a digest
is forging evidence about oneself.

\paragraph{And it is independently checkable by exactly that party.} A customer
who wishes to verify the claim can log raw parameters locally and recompute:
the digest is over the raw values, requires no knowledge of the tier table, and
the computation is public. The residual is invisible to \emph{us} and visible to
\emph{them}, which is the correct direction for a control the customer relies
on.

\paragraph{The threat model is unaffected.} What this system defends against is
an operator --- including us --- altering the record after it is written. The
chain's tamper-evidence properties~\cite{crosby2009} concern exactly that and
are untouched by who computed the digest.

The alternative is worth stating: for the server to compute the digest, the
server must receive the raw parameters. The residual is the price of the
property this paper is about, and there is no version of the design that has
both. \Cref{sec:residual} describes the principled way to buy it back.

\subsection{The action definition: the client is structurally prevented}\label{sec:definition}

An attested payload also records the \emph{definition} that governed the action:
which fields existed, which crossed, what a policy could have referenced. This
is what makes an old entry interpretable.

A client that could supply this field could attest that its action was governed
by a definition that never applied. That is worse than a wrong decision. It is
a false statement written into a signed, append-only ledger, and a wrong
decision at least shows up as a wrong decision, where this does not show up at
all. The forged definition would be
indistinguishable from the real one, permanently, and would corrupt exactly the
interpretive layer an auditor depends on.

So the definition is derived by the server from its own registry, and every
client-facing request schema is \emph{strict}: an unknown key is rejected, not
ignored. A client cannot supply the field because it cannot supply any field the
schema does not name.

Two details make this a property rather than an accident. First, strictness was
adopted for unrelated reasons, so nothing stated that this field in particular
must never be accepted --- a test now states it. Second, and more important, the
test's load-bearing half is its coverage check rather than the five schemas it
checks: the list of client-facing request bodies is derived
from the module's exports and compared against what is tested, so a sixth
request body added later fails the test until it is covered. The failure mode of
a hand-listed guard is the case nobody added to the list.

\subsection{The applied tier table: the client attests, the server compares}\label{sec:applied-tiers}

Between ``client is trusted'' and ``client is prevented'' lies the interesting
case.

The client applies a tier table to produce the minimized attributes. The server
knows which table it \emph{should} have applied. Neither can verify the other
directly: the server cannot see the raw parameters, and the client cannot see
the registry.

But both can compute a digest over the \emph{rules}. The client sends
$\mathrm{SHA\text{-}256}(\mathrm{JCS}(\textit{applied-tiers view}))$, a
canonical reduction of the definition it actually applied: each field's tier,
its normalized optionality, and its sorted projection names. The server computes
the same digest over the definition it authorized against and compares.

A mismatch means the two minimized under different rules: a stale client, a
registration that landed mid-flight, or a definition tampered with in transit.
The design decision we consider correct is that this is a fact about
disclosure, so it belongs in the chain rather than in a log line.

Two implementation details generalize.

\paragraph{The view is normalized, not hashed as given.}
\texttt{optional: undefined} and \texttt{optional: false} must not produce
different digests; projection names are sorted. Otherwise the two sides disagree
for reasons that have nothing to do with tiers.

\paragraph{The view is rebuilt rather than hashed as received.} The SDK holds the
wire shape, which may carry keys the contract does not cover: a description
field today, something else tomorrow. Hashing those would make the two sides
disagree about serialization rather than about disclosure. A digest intended to
detect one kind of divergence must be computed over exactly that kind and
nothing else, otherwise its false-positive rate destroys its value, and a
control that cries wolf gets turned off.

\subsection{The principle, generalized}

\begin{table}[h]
\centering\small
\begin{tabular}{@{}lll@{}}
\toprule
\textbf{Fact} & \textbf{Computed by} & \textbf{Why} \\
\midrule
Parameter digest   & Client, trusted   & Only party holding the data \\
Action definition  & Server only       & A client value would be undetectably false \\
Applied tier table & Client, \emph{compared} & Neither verifies the other's inputs; both share the rules \\
\bottomrule
\end{tabular}
\end{table}

The middle row is the one designs get wrong, because the field is
\emph{convenient} for the client to supply and the client already supplies its
neighbours. The test is not convenience or even trust in the client; it is what
a false value would look like afterward. \textbf{A fact whose falsification
would be undetectable in the permanent record must not be client-supplied at any
level of trust.}

\section{Disclosure is not authorization}\label{sec:disclosure}

\subsection{Deny-by-default protects one and not the other}\label{sec:deny}

Our mediator is deny-by-default: an action no policy permits is refused. This is
a complete protection for the authorization question. It is \emph{no}
protection at all for the disclosure question, and the asymmetry is easy to
miss because the same table drives both.

Re-tier a field from \tier{payload} to \tier{policy} and the raw value begins
crossing the boundary on every subsequent request, with no policy changed,
nothing redeployed, and no denial anywhere to signal it. In the policy corpus we
classified against, the highest-value sensitive field is
\texttt{send\_email.subject} --- free text, and referenced by none of the
policies in that corpus. A single re-tier of that one field starts protected content
flowing into the ledger on every send. Deny-by-default has nothing to
say about this, because nothing was denied. (The corpus is our own reference
policy set, not a customer's; we make no claim about the distribution in the
wild.)

Field tiering is a disclosure decision made with the same table as the
authorization decisions, and it needs controls of its own.

\subsection{The trust inversion}

The problem sharpens when the catalog becomes tenant-extensible. Before, the
tier table was code the customer shipped and reviewed, so a change required a
release. After, it is data the server sends, refreshed at runtime. An
administrator re-tiering a field in a console changes what every agent
discloses, immediately, with no code review anywhere in the loop.

This is a genuine regression in the customer's control, created by a feature
that is otherwise clearly right. Naming it is the first obligation, and the rest
of this section answers it.

\subsection{The narrow-only workload cap}\label{sec:floor}

The answer is a \emph{disclosure floor}: an optional, workload-side declaration
that caps what this workload will ever send, whatever the catalog says. The
catalog may narrow disclosure further; it may never widen past the floor.

\paragraph{The floor can only push a field \emph{down} to \tier{payload}.} There
is deliberately no way to raise one. A configuration that could widen disclosure
would be a second place to make exactly the mistake the floor exists to prevent,
and two mechanisms that can both widen are worse than one.

\paragraph{It is off by default}, and we think this is right despite being the
weaker security posture. A floor that had to be configured before anything
worked would make the extensible catalog useless, and a mechanism nobody enables
protects nobody. The trade is named in the design record rather than hidden.

The general shape --- \emph{the party bearing the risk holds a veto that can
only reduce exposure} --- is what we would reach for again.

\section{Leakage analysis}\label{sec:leak}

A projection is only as good as what it fails to remove. This section states,
per projection, what an observer of the minimized attributes learns. We use
``identifying'' in the ordinary re-identification sense~\cite{sweeney2002}, and
note that the standard caution about auxiliary information~\cite{narayanan2008}
applies to every row below.

\subsection{Per-projection}

\paragraph{\texttt{recipient\_domain}.} Discloses the recipient's organization.
For a large public provider this is near-zero information; for a small or
uniquely identifying domain it can approach identification of the counterparty,
though not of the individual. \emph{Residual:} the domain is a strong feature in
aggregate, since a sequence of decisions reveals an organization's
correspondents over time.

\paragraph{\texttt{payee\_known}.} One bit. Discloses whether a payee is on the
operator's allowlist and nothing about who they are. This is the strongest
projection in the set and the model for others: the policy question was always a
membership test, so the projection is exactly the predicate.

\paragraph{\texttt{endpoint\_host}.} Discloses which third-party service was
called. Frequently this \emph{is} the sensitive fact. Calling a specialist
medical API discloses category information about the subject even with every
identifier removed. We have no mitigation and do not claim one; a policy about
which services an agent may call cannot be enforced without knowing the service.

\paragraph{\texttt{endpoint\_path\_template}.} Discloses route structure, not
instances. \emph{Residual:} a route whose \emph{shape} is unique to a purpose
leaks that purpose, and segments that are semantically identifying but
syntactically ordinary (a username in a path segment, say) are preserved,
because the templating rule is syntactic. The honest statement is that the rule
catches identifier \emph{formats}, not identifier \emph{meanings}.

\paragraph{\texttt{resource\_path\_prefix}.} Discloses the directory, not the
leaf. \emph{Residual:} a directory hierarchy that encodes subject identity leaks
category, in that \texttt{/data/patients/\{id\}/} still tells you the record is a
patient record.

\subsection{The corpus, quantified}\label{sec:quant}

Our reference corpus classifies 19 fields across 5 action types:

\begin{table}[h]
\centering\small
\begin{tabular}{@{}lrr@{}}
\toprule
\textbf{Tier} & \textbf{Fields} & \textbf{Share} \\
\midrule
\tier{policy} (crosses raw)       & 11 & 58\% \\
\tier{derived} (crosses projected) &  6 & 32\% \\
\tier{payload} (never crosses)     &  2 & 11\% \\
\bottomrule
\end{tabular}
\end{table}

The two \tier{payload} fields are an email subject and a payment memo: both free
text, both referenced by no policy in the corpus.

The headline number is not the 11\%. Read carelessly, two-of-nineteen suggests
minimization barely does anything. The number that matters is that 8 of 19
fields, 42\%, never cross in their raw form, because the six \tier{derived}
fields are precisely the identifying ones: the recipient
address, the payee, the resource path, the external endpoint, the attachment
list. A per-class design would have had to choose between disclosing all six and
refusing every policy that depends on them.

The six \tier{derived} fields produce seven projected attributes, because one
field (the endpoint) projects to two independent policy-relevant abstractions:
its host and its route shape. That asymmetry is the argument
for making the projection a named function per target attribute rather than a
per-field transform.

What this quantification is and is not: it describes the corpus we classified
against --- our own reference policy set (\cref{sec:deny}) --- and is offered so
the shape of the classification is checkable, not as evidence about
the distribution in other organizations' policies. \Cref{sec:scope} states what
we can and cannot generalize.

\subsection{What the tier table itself discloses}

The applied-tiers digest (\cref{sec:applied-tiers}) is a hash over the
\emph{rules}, not the data. It discloses nothing about parameters. But the tier
table itself, once tenant-extensible, is a description of what an organization
considers sensitive --- mild but non-zero information, and it lives in our
registry.

\subsection{An incident: the projection that returned its input}\label{sec:incident}

Our first \texttt{resource\_path\_prefix} templated identifier-like segments
across the whole path and kept the leaf. It was tested and it looked right. It
failed on this repository's own sample resource identifier:
\texttt{patient-record-778812}.

Segment-wise templating left it completely intact, because the segment is not
wholly numeric and not a UUID: the identifier is embedded in a string with other
characters. The URL projection gets away with per-segment templating
only because URLs conventionally give an identifier its own segment; filesystem
resource identifiers do not follow that convention.

The leak guard described in \cref{sec:golden} caught it, specifically because it
searched the serialized output for the identifying value rather than checking
the field it expected the value in. The fix was to drop the final segment
entirely rather than attempt to sanitize it, on the reasoning that the leaf is
where the identifier lives, and a projection that tries to clean a string is a
projection whose correctness depends on enumerating formats.

What that costs, stated plainly: a policy can no longer discriminate on
a filename or extension. A rule reaching only \texttt{*.csv} files under
\texttt{/data/patients/} cannot be written; the rule can reach
\texttt{/data/patients/\{id\}/} and no further. We accepted it because filenames
are precisely where record identifiers and personal names appear (a file named
for its subject defeats every containment rule written about the directory
holding it), and because the containment questions customers actually ask us
are questions about directories: \emph{may this agent read outside the records
it was scoped to?} The extension was never the control; the location was.

We report this because it is the general lesson of the exercise: a projection
that attempts to remove sensitive substrings is fragile in a way that a
projection which discards a whole structural component is not. Prefer dropping a
component to cleaning one.

\section{Limits and residual trust}\label{sec:limits}

\subsection{Limits of the mechanism}

\paragraph{The mediator's decision rests on projections, so it can be wrong in
ways the raw path would not be.} A policy that would have matched on a full path
may not match on a directory. This is visible at authoring time --- the
attribute simply does not exist in the schema --- which is the design's answer,
but an answer that depends on operators reading schemas.

\paragraph{Nothing here defends against a compromised workload.} An agent whose
process is controlled by an attacker discloses whatever it likes over its own
network connections; minimization governs what crosses \emph{our} boundary, not
every boundary.

\paragraph{The tier table is a static, per-field, per-action classification.} It
cannot express context-dependence (\emph{this field is sensitive when the
recipient is external}), and we have not needed it. We expect that need to
arrive, and the design has no answer for it today.

\subsection{Residual trust, and the principled way to remove it}\label{sec:residual}

The ledger commits to the client's claim about its own parameters
(\cref{sec:digest-trust}). An auditor can confirm a candidate parameter set
matches the digest; they cannot establish that the client did not compute the
digest over something else. The mediator cannot check this, by construction.
What the design does provide is that the \emph{disclosure rules} under which the
client operated are attested and compared (\cref{sec:applied-tiers}), and that
the \emph{definition} is beyond the client's reach (\cref{sec:definition}).

We think the principled way to close this is known and we have not built it.
What is wanted is a proof, verifiable by the mediator, that the minimized
attributes are a correct application of a stated tier table to \emph{some}
preimage of the committed digest, a zero-knowledge proof of correct
projection~\cite{gmr1989,parno2013}. That would make the digest and the
attributes jointly checkable without the mediator learning the parameters, which
is precisely the property \cref{sec:digest} obtains structurally and cannot
obtain cryptographically. The obstacles are practical rather than theoretical:
the projections include string operations that are expensive to express as
arithmetic circuits, and the mediation path is latency-sensitive. We regard this
as the natural next piece of work on this design and the place where the
structural approach of this paper meets the cryptographic one.

\subsection{Deployment scope, and what it means for these claims}\label{sec:scope}

The system is deployed as a small number of private pilots, not as a general
release. This is a deliberate choice rather than a stage we are passing through:
what a governance layer must express is determined by what organizations
actually need governed, and that is learned by working closely with a few
deployments rather than broadly across many. A control plane that reaches
general availability before its designers understand the requirement tends to
have made the requirement up.

\paragraph{What we do claim.} The design is implemented, running, and exercised
by tests that pin every tier assignment against hand-written literals and
compare digests across two language implementations. The failures reported in
\cref{sec:drift} and \cref{sec:incident} are real and were found by these
mechanisms. The properties argued in \cref{sec:digest} and
\cref{sec:asymmetry} are structural --- they follow from where the commitment
is computed and which party can supply which field --- and do not depend on
deployment scale.

\paragraph{What we do not claim.} We have not shown that the three tiers are
sufficient for policy corpora other than the one we classified against, nor that
our five projections cover the space of policy-relevant abstractions, nor
anything about operator behaviour at scale, in particular whether the
disclosure controls of \cref{sec:disclosure} are ones administrators actually
reach for. The floor of \cref{sec:floor} is off by default, and we have no
evidence about how often it is enabled, because there is not yet a population to
measure. These are questions about adoption, and they need a larger and more
varied deployment base than we have.

\subsection{How the design arrived at its present shape}

Because a design presented as a finished argument can look as though it was
reasoned out in advance, we record what actually happened. The system has been
under development since May 2026, and the minimization design in this paper has
been through four generations. Each was forced by a specific finding rather
than by a change of taste.

\begin{table}[h]
\centering\small
\begin{tabular}{@{}p{4mm}p{58mm}p{62mm}@{}}
\toprule
& \textbf{Change} & \textbf{What forced it} \\
\midrule
1 & A hand-written per-field tier table, a Cedar schema derived from it, and client-side digest builders in two languages & The per-class alternative had no good setting (\cref{sec:field-not-class}) \\
2 & Resource paths project to a \emph{directory} rather than a templated full path & The leak in \cref{sec:incident}: an identifier survived segment-wise templating intact \\
3 & Tier table, policy schema and wire schema \emph{generated} from one declaration; the equivalence test guarding the generator replaced by golden literals & Three hand-maintained artifacts had already drifted once (\cref{sec:drift}), and the generator's own safety argument had gone vacuous (\cref{sec:golden}) \\
4 & The client \emph{attests the tier table it applied}; the server compares (\cref{sec:applied-tiers}) & Asking whether the client and the server could disagree about a definition, and finding that they could not, because the client had none \\
\bottomrule
\end{tabular}
\end{table}

Two of these four were reversals of something we had already reviewed and
shipped, and one --- the fourth --- is the paper's strongest contribution and
was not in the original design at all. It arrived from a question rather than from a
plan.

We report this because the asymmetry principle of \cref{sec:asymmetry} in
particular reads as though it were derived top-down. It was not. It was
extracted from three decisions taken separately, for local reasons, over several
weeks, and the principle became visible only once they sat beside each other.

\section{Related work}\label{sec:related}

\paragraph{Information-flow control and declassification.} The tier vocabulary
is a lattice~\cite{denning1976} and the \tier{derived} projections are
declassifiers, which places this work in the tradition of language-based
information-flow security~\cite{sabelfeld2003,myers1997}. Our setting is
narrower and the mechanism correspondingly cruder: classification is per-field
and static rather than derived from program analysis, and enforcement is a
library boundary rather than a type system. What we add to the picture is the
\emph{attestation} of which declassification policy was
applied~(\cref{sec:applied-tiers}), which is not usually a concern in that
literature~\cite{sabelfeld2009} because the policy is compiled in.

\paragraph{Data minimization in regulation and practice.} The minimization
principle is a legal requirement~\cite{gdpr} without an implementation, and the
engineering literature on privacy by design~\cite{gurses2011,pfitzmann2010}
supplies goals more readily than mechanisms. Most engineering responses operate
at the storage layer: redaction, tokenization, field-level encryption. The
distinguishing feature here is that minimization occurs \emph{before} the trust
boundary rather than at rest, so the data does not need protecting because it
does not arrive.

\paragraph{Accountability logging and redaction.} Designing logs to evidence
compliance while limiting what they retain is a recognised
problem~\cite{butin2013,butin2014}, and structured-logging practice typically
classifies at the log-statement level, the class-level design we rejected in
\cref{sec:field-not-class}. Redaction at write time additionally breaks the
commitment property of \cref{sec:digest}: an entry then commits to a redacted
view, and a reader cannot tell what was removed.

The cryptographic form of this idea is closer to our concerns and deserves an
explicit comparison. \emph{Redactable} and \emph{sanitizable} signature
schemes~\cite{steinfeld2001,johnson2002,ateniese2005} allow a signed document to
have parts removed while the remainder still verifies, which sounds like exactly
what an evidence package wants. The difference is where the sensitive value goes.
Those schemes let a holder redact a record \emph{after} signing, so the data must
first exist in signed form: it has crossed the mediation and ledger boundaries
of \cref{sec:boundaries} and been committed to by a signer who saw it. Our
concern is upstream of that: the identifying value never reaches the signer at
all, and what is committed is a digest computed where the data already lives.
The two are complementary rather than competing --- redactable signatures govern
what a holder may disclose from a record, and this design governs what enters the
record --- and a system wanting both would apply them at different points.

\paragraph{Transparency logs and certificate transparency} supply the
append-only, externally verifiable ledger model~\cite{rfc6962,crosby2009}. To
our knowledge they do not address the tension between an immutable record and a
mutable disclosure policy, because the logged objects are public by
construction.

\paragraph{Agent governance, and why this is a runtime problem.} Work on
governing autonomous agents at runtime is early but no longer absent: a recent
survey maps the agent-identity standards landscape and its
gaps~\cite{otsuka2026}, and composition and provenance for agentic systems are
receiving dedicated treatment~\cite{agentriskbom}. We are not aware of prior
work on the disclosure question this paper addresses. The nearest adjacent work
is build-time: static
validation can enforce agent patterns and catch classes of defect before code
ships~\cite{agentverifier}. It cannot constrain what a probabilistic agent
chooses to do with parameters it composes at runtime, which is the case
addressed here: an agent's action is selected by a model rather than programmed,
so the parameters that must be authorized and audited do not exist until the
moment of the call. That runtime framing, and the decomposition of the
problem into which this mechanism fits, are set out in~\cite{c1primitives}. The agent framework this governance layer was
built alongside~\cite{auriteframework} is representative of that class. We note
the distinction rather than claim the ground: build-time validation and runtime
mediation are complements, and a deployment wanting assurance needs both.

\paragraph{Privacy-preserving authorization} --- attribute-based and anonymous
credentials~\cite{chaum1985,camenisch2001}, and zero-knowledge proofs of policy
satisfaction~\cite{gmr1989} --- solves a stronger version of this problem with
cryptographic guarantees rather than structural ones. The comparison is
unflattering to us in rigour and flattering in deployability: our projections
are ordinary functions any auditor can read, run and reason about without a
trusted setup, and the whole system is a library change rather than a protocol
change. We regard \cref{sec:leak}'s honest residuals as the price of that.

\section{Conclusion}

The mediation point of an agent governance system is where the pressure to see
everything and the pressure to record everything meet, and it is also the point
where those pressures are most easily mistaken for one requirement. They are
not. A policy engine needs predicates; a ledger needs a commitment; neither
needs values, and the values that must cross for policy reasons are usually not
the identifying ones.

Separating them turns out to hinge on one placement decision. Compute the
commitment over the raw parameters, before minimization, and disclosure policy
becomes mutable over an immutable record: tiers can be reclassified,
minimization can be retrofitted to a populated chain, and an already-shipped
offline verifier never learns that any of it happened. Compute it over the
minimized view and none of that is available.

The second lesson was less expected. Minimization redistributes \emph{who knows
what}, and therefore forces a question about who may attest to what. The answer
we arrived at was derived from a specific near-miss: the right party is the one
who cannot lie undetectably, and a fact failing that test must be structurally
unavailable to the client rather than merely discouraged. We suspect it
generalizes to any system that both minimizes and attests.

The last lesson is the one we would tell a team starting this work: \textbf{the
tier table is a disclosure decision, and deny-by-default does not protect it.}
Every control we built for the authorization half was already there and worked.
Every control for the disclosure half had to be built separately, and the fact
that one table drives both is exactly why it took us an incident to notice.

\section*{Data availability}

A reference implementation is in preparation and will be released under the MIT
licence: the five projections of \cref{sec:projections}, the per-action tier
declaration format of \cref{sec:generation}, and the applied-tiers digest of
\cref{sec:applied-tiers}. The purpose of releasing it is that
\cref{sec:leak}'s leakage claims should be checkable rather than believed: each
projection is a pure function of a few lines, and a reader who disagrees with
our analysis can run it. It is not available at the time of writing.

It is intended as a frozen artifact accompanying this paper, not a supported
library. We expect the projection set to grow in our own system and make no
undertaking to keep the published version in step.

The mediator's internals, the policy corpus, and the catalog registry API are
deliberately not included. They are not part of the verification surface:
nothing in \cref{sec:leak} depends on them, and a reader cannot check any claim
in this paper more thoroughly by having them.

\appendix
\section{A request, end to end}\label{app:worked}

Every value below is taken from the golden corpus described in
\cref{sec:golden}, which pins them as literals in the test suite. Nothing here
is illustrative.

\subsection{\texttt{send\_email}}

The agent calls the SDK with:

\begin{lstlisting}
{
  "action_type": "send_email",
  "recipient": "Jane.Doe@Customer.COM",
  "subject": "patient 778812 discharge summary",
  "body_size_bytes": 2048,
  "attachment_hashes": ["aaaa...(64 hex)", "bbbb...(64 hex)"]
}
\end{lstlisting}

\paragraph{Step 1, commit over the raw parameters.} The SDK computes
$\mathrm{SHA\text{-}256}(\mathrm{JCS}(\textit{parameters}))$ over the object
exactly as given, before anything else happens. This is the value the ledger
will carry, and \cref{sec:digest}'s whole argument is that it is fixed at this
moment, independent of every classification decision below.

\paragraph{Step 2, minimize.} The tier table for \texttt{send\_email}
classifies four fields: \texttt{recipient} (\tier{derived}),
\texttt{body\_size\_bytes} (\tier{policy}), \texttt{attachment\_hashes}
(\tier{derived}), \texttt{subject} (\tier{payload}). What crosses the workload
boundary is:

\begin{lstlisting}
{
  "recipient_domain": "customer.com",
  "body_size_bytes": 2048,
  "attachment_count": 2
}
\end{lstlisting}

Three things happened. The recipient was projected to its domain, lowercased,
so \texttt{Jane.Doe} did not cross. The attachment \emph{hashes} became a
\emph{count}, since no policy in the corpus asks which files, only how many. And
the subject --- \emph{``patient 778812 discharge summary''}, the highest-value
sensitive value in the whole request --- did not cross at all, and does not
appear in the ledger except as an input to the Step 1 digest.

The corresponding test asserts that the strings \texttt{Jane.Doe},
\texttt{patient 778812 discharge summary} and \texttt{778812} appear nowhere in
the serialized output, under any key (\cref{sec:golden}).

\paragraph{Step 3, attest the disclosure rules.} The SDK computes the
applied-tiers digest over the tier table it used and sends it alongside
(\cref{sec:applied-tiers}). The server computes the same digest over the
definition it authorized against and compares.

\paragraph{Step 4, evaluate.} Cedar sees only the three attributes above. A
policy \texttt{recipient\_domain == "customer.com"} matches; a policy about the
subject line cannot be written, because that attribute does not exist in the
generated schema, which is the design's answer to silent non-matching
(\cref{sec:constraints}).

\paragraph{Step 5, attest.} The event records the action type, the decision,
the matched policy, the applied-tiers digest, and the Step 1 hash.
An auditor holding a candidate parameter set can confirm it is the one,
exactly, including the subject line that never left the workload.

\subsection{The projections on harder input}

\begin{table}[h]
\centering\footnotesize
\begin{tabular}{@{}p{54mm}p{32mm}p{42mm}@{}}
\toprule
\textbf{Raw} & \textbf{Projected} & \textbf{Note} \\
\midrule
\texttt{/data/patients/}\newline\texttt{40b1c2d3-.../rec.json} & \texttt{/data/patients/\{id\}/} & UUID templated, leaf dropped \\
\texttt{patient-record-778812} & \emph{attribute omitted} & No directory at all \\
\texttt{nobody-at-example} (no \texttt{@}) & \texttt{""} & Unparseable $\rightarrow$ matches no domain policy \\
\bottomrule
\end{tabular}
\end{table}

The second row is worth dwelling on. A resource with no directory emits no
attribute, rather than an empty one. Absent beats empty: Cedar's
\texttt{has} then means what it says, and a policy written \texttt{like "*"}
cannot quietly match a resource whose directory was never known. It is also the
input that defeated the first implementation (\cref{sec:incident}).


\end{document}